\documentclass[modern]{aastex63}

\hypersetup{linkcolor=red,citecolor=blue,filecolor=cyan,urlcolor=magenta}

\newcommand\mum{$\mu$m}
\newcommand\ra{$\rightarrow$}

\received{}
\revised{}
\accepted{22-Dec-2021}
\submitjournal{ApJ}

\shorttitle{R Aqr SOFIA/FORCAST Observations II}
\shortauthors{Sankrit et al.}

\begin{document}

\title{SOFIA/FORCAST Monitoring of the Dust Emission from R~Aqr: Start of the Eclipse}

\correspondingauthor{Ravi Sankrit}
\email{rsankrit@stsci.edu}

\author[0000-0001-8858-1943]{Ravi Sankrit}
\affiliation{Space Telescope Science Institute, 
3700 San Martin Drive, Baltimore, MD 21218, USA}

\author[0000-0003-1878-8310]{Eric Omelian}
\affiliation{Space Science Institute, 
4765 Walnut St., Suite B, Boulder, CO 80301, USA}

\author[0000-0002-3311-5918]{Uma Gorti}
\affiliation{Carl Sagan Center, SETI Institute, Mountain View, CA 94043, USA} 
\affiliation{NASA Ames Research Center, Moffett Field, CA 94035, USA}

\author[0000-0003-1892-2751]{R.\ Mark Wagner}
\affiliation{The Ohio State University, LBT Observatory, 
933 N. Cherry Ave, Tucson, AZ 85721, USA}

\author[0000-0002-8937-3844]{Steven Goldman}
\affiliation{Space Telescope Science Institute, 
3700 San Martin Drive, Baltimore, MD 21218, USA}

\author[0000-0002-4678-4432]{Patricia A. Whitelock}
\affiliation{South African Astronomical Observatory, 
P.O. Box 9, Observatory, 7935 Observatory, South Africa}
\affiliation{Department of Astronomy, University of Cape Town, 
Private Bag X3, Rondebosch 7701, South Africa}

\begin{abstract}
We present mid-infrared spectra from our continued monitoring of
R~Aquarii, the nearest symbiotic Mira, using the Stratospheric
Observatory for Infrared Astronomy (SOFIA).  New photometric and
spectroscipic data were obtained with the Faint Object infraRed
CAmera for the SOFIA Telescope (FORCAST) in 2018 and 2019 after the
system had started its ``eclipse'', during which it became two
magnitudes fainter in the visual.  The mid-IR flux, in particular
the 10\mum\ silicate feature, have strengthened compared with the
previous cycles.  Radiative transfer models for the circumstellar
dust emission were calculated for the new spectra, and re-calculated
for those previously obtained using more appropriate values of the
near-IR magnitudes to constrain the properties of the AGB spectra
heating the dust.  The modeling shows that the luminosity dependence
on pulsation phase is not affected by the onset of the eclipse, and
that the increase in the mid-IR flux is due to a higher dust density.
The models also confirm our earlier results that micron-size grains
are present, and that no changes in the grain composition are
required to explain the variations in the spectra.
\end{abstract}

\keywords{Asymptotic giant branch (108); 
          Symbiotic binary stars (1674); 
          Circumstellar dust (236); 
          Stellar pulsations (1625)}

\section{Introduction} 

R~Aquarii (R~Aqr), at a distance of 218~pc \citep{min14} is the
closest symbiotic system.  It consists of an M-type Mira variable
with a steady pulsation period of 387 days \citep{belczynski00},
and a hot compact companion, which has a temperature of about
60,000~K, and a luminosity in the range 5--20~$L_{\odot}$ \citep{kaler81,
burgarella92}.  The hot star is either a White Dwarf (WD) or a
subdwarf on the evolutionary track toward a WD state \citep{burgarella92}.
Henceforth we will use the term WD for the hot component of the
binary.  The WD sustains a compact ionized nebula \citep{meier95}
beyond which lie complex nebular structures extending several
arcminutes away from the central binary, which are probably due to
nova-like ejections of material many centuries ago \citep{solf85,
liimets18}.  The system also contains a spectacular jet
\citep[e.g.][]{melnikov18} that is most likely driven by the accretion
flow of material from the cool Asymptotic Giant Branch (AGB) star
onto the WD\@.  R~Aqr is a ``Dusty'', or D-type symbiotic, because
it is a bright infra-red (IR) source. It is also known as a symbiotic
Mira \citep{whitelock87}, although  it is rather less dusty than
is typical for this class. The IR emission is dominated by the
circumstellar shell, and the spectral energy distribution (SED)
consists of the stellar blackbody plus a thermal dust component.
An early spectrum \citep{merrill76} revealed the presence of the
prominent 9.7\,\mum\ silicate feature, known to be characteristic
of M-type AGB stars \citep{woolf73}.

Based on successive periods of visual dimming of R~Aqr during
1928--1934 and 1974--1980, \citet{willson81} inferred an orbital
period of about 44 years, and suggested that the decrease in
brightness was due to a dusty accretion stream obscuring the Mira.
\citet{gromadzki09a} refined the period to 43.6~years, and showed
that the dimming occurs as the WD eclipses its Mira companion, and
that the periastron passages occur during these times.  The current
eclipse has started (\S2.2, Fig.~\ref{figlc}; note that henceforth
we will use the term ``eclipse'' to refer generally to this period
of visual dimming), and the periastron will occur in 2023.

We initiated a program in 2016 to monitor R~Aqr at mid-IR wavelengths
using the Stratospheric Observatory for Infrared Astronomy (SOFIA).
In \citet{omelian20} (Paper 1) we presented observations using
SOFIA/FORCAST in 2016, and 2017, supplemented with observations
using the Infrared Space Observatory (ISO) obtained in 1996, when
the system was near apastron.  We modeled the mid-IR SED at these
epochs and, based on the strengths of the silicate features, showed
that there had been a significant decrease in the dust density
between apastron and the current epoch.  The models also showed
that the change in the SED between 2016 and 2017 did not require
any modifications to the composition of the emitting grains, but
instead could be explained by variations with pulsation-phase of
the effective temperature of the AGB star, the inner radius of the
dust shell, and the dust density at this inner radius.

In this paper we present the SOFIA/FORCAST observations of R~Aqr
obtained during Cycle 6 (2018) and Cycle 7 (2019), and follow the
evolution of the mid-IR SED.  The new data are modeled using the
method described in Paper 1.  We also calculate revised models for
the ISO, and SOFIA Cycles 4 and 5 data using updated values for the
near-IR ($JHKL$) fluxes.

The observations are presented in \S2, and the FORCAST flux
measurements in \S3.  In \S4 the modeling procedure is outlined,
and the need for updated models for the earlier data is explained.
The results and their implications are discussed in \S5, and \S6
contains some concluding remarks.


\section{Observations}

As part of our continued monitoring of R~Aqr in the mid-IR, we
obtained photometric and spectroscopic data using FORCAST on board
SOFIA \citep{young12} on August 24, 2018 and October 16, 2019 (all
dates are UTC) during Observing Cycles 6 and 7, respectively.  The
FORCAST instrument \citep{herter13} has a suite of imaging filters
and low-dispersion grisms, and covers the 5.4--37.5\,\mum\ wavelength
region, except for a gap between about 13.5 and 17.5\,\mum.  Although
this gap wavelength region has been accessed on airborne observatories
\citep[e.g.][]{forrest79}, the strong molecular absorption bands
reduce the transmission close to zero even in the stratosphere.

These new observations were obtained using the same strategy as in
Cycles 4 and 5 (Paper 1).  All the standard photometric filters
between 6.4\,\mum\ and 37.5\,\mum\ were used in dual-channel mode.
Although the overall wavelength coverage remains the same, the set
of filters is not identical from cycle-to-cycle.  Spectroscopic
observations were obtained using the G111 (8.4--13.7\,\mum) and
G227 (17.6--27.7\,\mum) grisms in both cycles.  Additionally, in
Cycle 7, spectra were obtained  using the G329 (28.7--37.1\,\mum)
grism.  All spectra were taken through the 4.7\arcsec\ slit yielding
spectral resolutions, R = 130, 110 and 160 for the G111, G227, and
G329 data, respectively.  The data were obtained using standard
chopping and nodding, along with 3-point dithers to remove the
effects of bad pixels.  The exposure times were dominated by overhead,
with on-source times of approximately 30 to 60 seconds for each
setting.

For this study, we used the flux-calibrated ``Level 3'' data products
that are provided by the SOFIA Science Center.  The flux calibration
factors are obtained from observations of standard stars, except
in the case of G227 and G329 spectra, which use bright asteroids.
The details are available in the ``Guest Observer Handbook for
FORCAST Data Products'' provided by SOFIA, and available on their
website.  The wavelength coverage of the G111 grism includes the
ozone feature at 9.6\,\mum\ over which the calibration is unreliable,
and therefore we exclude the spectral data between 9.2 and 10.3\,\mum\
in our analyses.  The long-wavelength end of the G329 spectrum is
poorly calibrated due to a strong terrestrial feature, and therefore
we exclude the data longward of 35\,\mum.

\begin{deluxetable*}{clcc}[ht!]
\tablecaption{Mid-IR Observations of R~Aqr \label{tabobs}}
\tablewidth{0pt}
\tablehead{
\colhead{Observation} 
& \colhead{Obs.\ Date}
& \colhead{V\tablenotemark{a}}
& \colhead{$\phi_{Mira}$\tablenotemark{b}}
}
\startdata
ISO & 1996 May 16 & 8.0 & 0.14 \\
SFC4\tablenotemark{c} & 2016 Sep 22 & 10.6 & 0.35\\
SFC5 & 2017 Aug 7 & 8.5 & 0.17 \\
SFC6 & 2018 Aug 24 & 9.4 & 0.16 \\
SFC7 & 2019 Oct 16 & 11.5 & 0.24 \\
\enddata
\tablenotetext{a}{Visual magnitude from AAVSO}
\tablenotetext{b}{$\phi_{Mira}=0.0$ (visual maximum) on 2005 October 6, 
period = 387.0 days}
\tablenotetext{c}{SOFIA Cycle 4, et.\ seq.}

\end{deluxetable*}

\begin{figure}[ht!]
\plotone{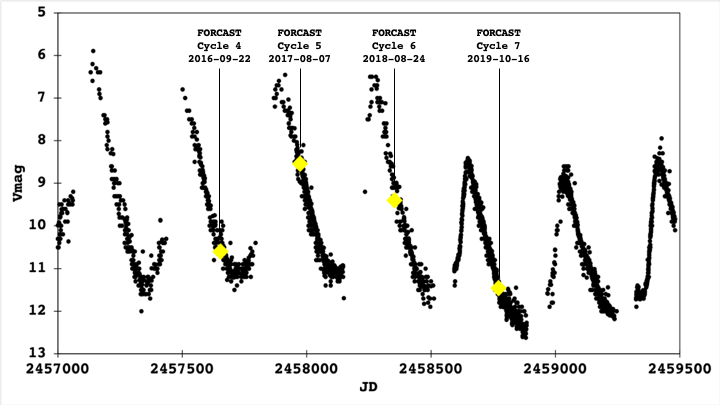}
\caption{The AAVSO light curve of R Aqr from December 2014 through
September 2021. Yellow diamonds indicate the SOFIA FORCAST observations
to date.}
\label{figlc}
\end{figure}

Table~\ref{tabobs} lists the mid-IR observations, the dates they
were obtained, and the visual magnitude and Mira phase on the
observation date.   Figure \ref{figlc} shows the light curve for R
Aqr from December 2014 through September 2021.  The system was about
two magnitudes fainter in the visual at the three most recent
pulsation maxima (2019 mid-June, 2020 early-July, and 2021 late-July)
compared with previous cycles, indicating that the eclipse has
started.  The Cycle 4 and 5 SOFIA/FORCAST observations were obtained
before the onset of the eclipse, and Cycle 7 observations after it
started (Fig.~\ref{figlc}).  The Cycles 5 and 6 observations were
obtained at approximately the same Mira pulsation phase, but the
system was about one magnitude fainter on the latter date than on
the former (Table~\ref{tabobs}).  This suggests that the dimming
had started before the Cycle 6 observations were obtained.

\subsection{Historical JHKL Data}

R~Aqr was monitored at near-IR wavelengths between 1980 and 2014
using the MkII photometer on the 0.75m telescope at the Sutherland
site of the South African Astronomical Observatory.  The early data
were published and/or discussed by \citet{catchpole79, whitelock83,
whitelock87, gromadzki09b}, but the entire data-set is presented
in Table \ref{tabIRphot} as the monitoring program was terminated
in 2014.  These data are on the photometric system described by
\citet{carter90}, which can be converted to the 2MASS system following
\citet{carpenter01}. This involved some very small ($<0.05$ mag)
corrections to the data published prior to 1990. The photometry is
accurate to better than $\pm 0.03$ in the $JHK$ bands, and $\pm
0.05$ in the $L$ band.

We use these data to select the best representative values for the
dates of the FORCAST observations based on the Mira pulsation phase
and visual magnitude.  This is discussed further in \S4.

\section{Results: the Mid-IR Spectrum of R Aqr}

As in the previous data, the photometric images obtained in Cycles
6 and 7 show that at these wavelengths, R~Aqr is consistent with a
point source at the FORCAST angular resolution, which ranges from
just under 3\arcsec\ at 5.4\,\mum\ to about 5\arcsec\ at 37\,\mum.
The mid-infrared emission is thus dominated by the cool AGB star,
its accompanying circumstellar dust shell, and possibly the accretion
flow.  There is negligible contribution from the surrounding nebula,
which is prominent at optical wavelengths \citep[e.g.][]{liimets18}.

The fluxes were extracted from the images using aperture photometry
as in Paper 1, following the SOFIA FORCAST Photometry Cookbook
(version of 4 Sep 2018, Rev A).   The source flux is extracted from
a 12 pixel radius circular aperture, centered on the flux-weighted
centroid of the target.  The source radius of 12 pixels corresponds
to 9.216\arcsec\@.  The background is defined by an annulus between
radii of 15 and 25 pixels centered at the same point as the target.
Since the data are already sky-subtracted by the pipeline, the
background is used only to establish the photometric uncertainty.
In all the images, the median fluxes in the background regions were
identically zero, and the standard deviations ranged between 0.01
and 0.03~Jy. The source is sufficiently bright and the photometric
uncertainties, $\eta_{\rm{m}}$, range from 0.05\% to 0.63\%.  The
uncertainties in the flux are dominated by the relative error in
flux calibration, $\eta_{\rm{flux}}$, obtained from the FITS header
values, CALFCTR and ERRCALF, for each observation. These uncertainties
range from 1.89 to 2.88\% in the SOFIA Cycle 6 data, and 0.90 to
6.10\% in the SOFIA Cycle 7 data.  The Cookbook lists a third source
of uncertainty, $\eta_{\rm{model}} \approx 0.05$, due to the flux
calibration model.  Since our data in each epoch were obtained on
the same flight-legs, the relative uncertainty from the model should
be significantly lower, and therefore we do not include this term
in our error calculation.  However, we assume that the minimum
combined uncertainty for each measurement is 5\%.

\begin{deluxetable*}{ccccccc}[ht!]
\tablecaption{R Aqr: SOFIA FORCAST Photometry \label{tabphot}}
\tablewidth{0pt}
\tablehead{
\colhead{Filter} 
& \colhead{$\lambda_{eff}$}
& \colhead{$\Delta$$\lambda$}
& \colhead{Flux SFC4}
& \colhead{Flux SFC5} 
& \colhead{Flux SFC6}
& \colhead{Flux SFC7} \\
\colhead{}       
& \colhead{(\mum)}
& \colhead{(\mum)}
& \colhead{(Jy)}
& \colhead{(Jy)}
& \colhead{(Jy)}
& \colhead{(Jy)}                 
}
\startdata
F056  & 5.6   & 0.08  & ...  & 1583 ($\pm$380) & ... & ... \\ 
F064  & 6.4   & 0.14  & 747 ($\pm$37)  & 1149 ($\pm$58) & 1248 ($\pm$62) & 974 ($\pm$49) \\ 
F077  & 7.7   & 0.47  & 580 ($\pm$29)  & 835 ($\pm$42) & 963 ($\pm$48) & 828 ($\pm$43) \\
F088  & 8.8 & 0.41 & ... & ... & 1028 ($\pm$51)  & ... \\  
F111  & 11.1  & 0.95  & 661 ($\pm$79) & 883 ($\pm$44) & 1024 ($\pm$51) & 851 ($\pm$52) \\ 
F197  & 19.7  & 5.5   & 359 ($\pm$18) & 435 ($\pm$22) & ... & 441 ($\pm$22) \\ 
F253  & 25.3  & 1.86  & 203 ($\pm$16) & 254 ($\pm$13) & 275 ($\pm$14) & 251 ($\pm$13) \\ 
F315  & 31.5  & 5.7   & 143 ($\pm$7) & 195 ($\pm$14) & 204 ($\pm$10) & 180 ($\pm$9) \\ 
F336  & 33.6  & 1.9   & 131 ($\pm$17) & 163 ($\pm$8) & 181 ($\pm$9) & 161 ($\pm$8) \\ 
F348  & 34.8  & 3.8   & 110 ($\pm$6) & 139 ($\pm$15) & 147 ($\pm$7) & 148 ($\pm$7) \\ 
F371  & 37.1  & 3.3   & 92 ($\pm$6)  & 121 ($\pm$7) & 129 ($\pm$7) & 116 ($\pm$6) \\
\enddata

\end{deluxetable*}

The grism data are subject to slit-losses that are variable, which
adds uncertainty to the absolute flux calibration since these losses
may be different for the science target and calibrator.  Therefore,
we scale the spectra to photometric points, using the 11.1\,\mum\
flux for the G111 spectra, and the 19.7\,\mum\ flux for the G227
spectra. The 19.7\,\mum\ data were not obtained during Cycle 6, and
the G227 spectrum was scaled instead to the 25.3\,\mum\ photometric
flux.  Scaling the G111 spectra by convolving with the throughput
curve of the SOFIA FORCAST 11.1\,\mum\ filter yields a difference
of less than 0.5\%.  The G227 grism bandpass does not completely
overlap the SOFIA FORCAST 19.7\,\mum\ filter throughput.  However
the spectrum is relatively smooth in that wavelength region, and
we expect the difference between the scaling at one wavelength point
and convolving with the filter throughput would yield a negligible
difference as for the shorter wavelength filter. The G329 grism
(28.7-37.1 $\mu$m) is also scaled to a photometric point at 33.6
$\mu$m by convolving the G329 spectrum with the throughput filter
of the 33.6 $\mu$m filter. The relative errors in the spectral
fluxes for a single spectrum (outside the region affected by ozone
for G111 data) are approximately 2\%, as determined from the median
of the error arrays.

\begin{figure}[ht!]
\plotone{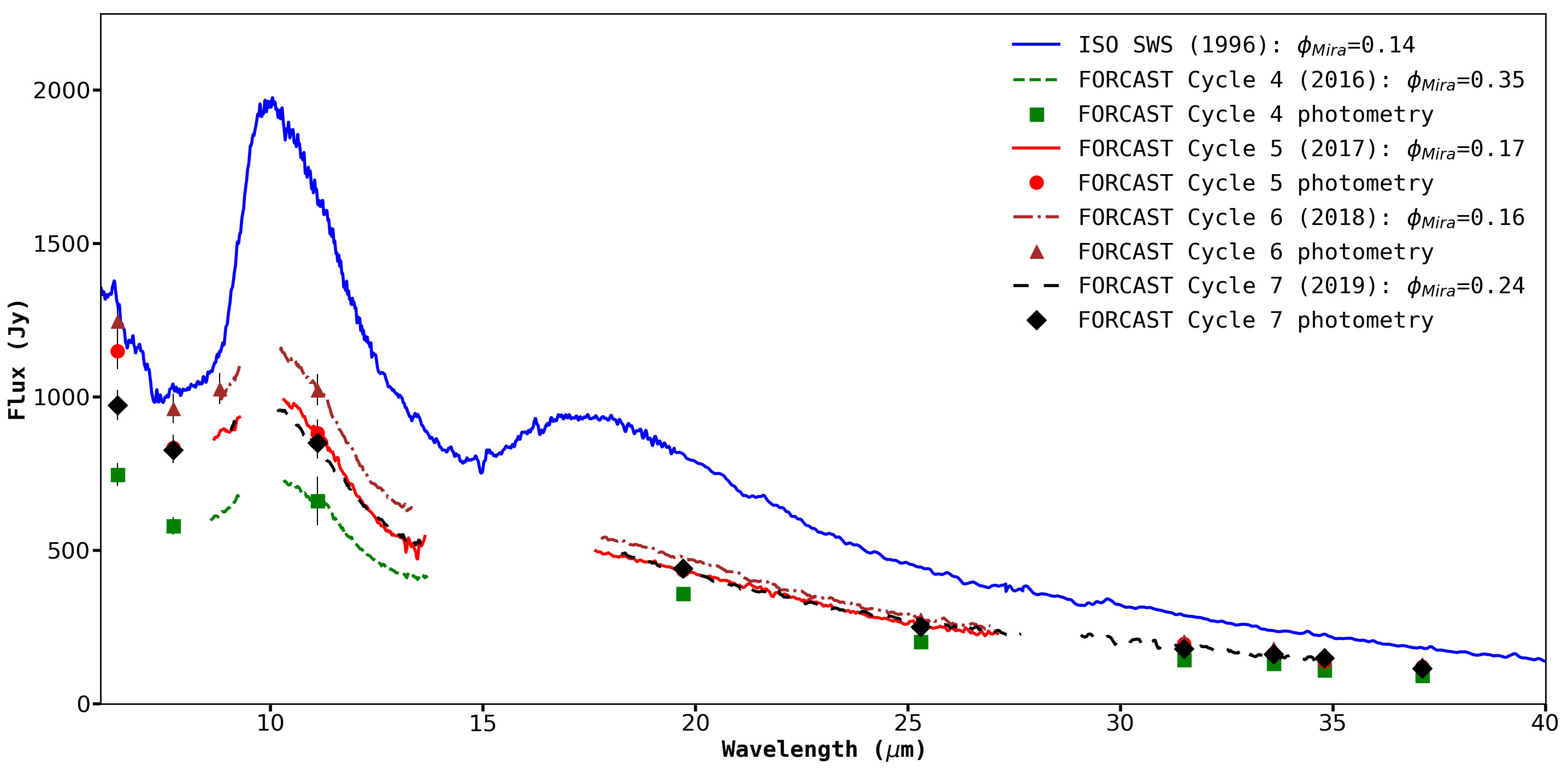}
\caption{Infrared Spectral Energy Distribution (SED) of R Aqr
obtained with ISO, and with SOFIA FORCAST. The spectra around 10
$\mu$m for the SOFIA FORCAST grism data has been removed due to the
telluric atmospheric ozone feature that occurs around this wavelength.
The error-bars on the photometric measurements are shown, and are
mostly smaller than the symbol size.}
\label{figdata}
\end{figure}

Table~\ref{tabphot} lists the filter names, central wavelengths and
bandpasses, and measured fluxes and errors for the imaging data
from all observing cycles. Figure~\ref{figdata} shows plots of the
photometric points and the grism spectra, and also includes the
ISO-SWS spectrum obtained in 1996.  The classic silicate spectral
features around 10 and 18\,\mum\ are evident in the spectral energy
distributions at each epoch.

\section{Radiative Transfer Models of the Dust Emission}

In an oxygen-rich AGB star, silicate grains responsible for the 10
and 18\,\mum\ features condense in a region of the circumstellar
shell just beyond the molecular layer, typically at a distance of
about 5--10 stellar radii from the center \citep[e.g.][]{gobrecht16}.
The silicate features exhibit a wide range of shapes and relative
strengths \citep{speck00, kraemer02}, which depend on the density
and temperature of the dust, and on the composition of the grains
and their size distribution.

We model the emission from the dust shell using the publicly available
three-dimensional radiative transfer code, RADMC-3D \citep{dullemond12},
using the same procedure as in Paper 1.  Briefly, the AGB star is
placed at the center of the grid, and the dust shell is spherically
symmetric around the star, extending between an inner radius $R_{in}$
and an outer radius $R_{out} = 10,000\,\rm{au}$.  The radial dust
density profile is assumed to be $\rho_{dust}(r) =
\rho_{in}(r/R_{in})^{-2}$, and is specified at 100 logarithmically
spaced intervals in $r$.

The optical constants used are those for oxygen-deficient circumstellar
silicate dust from \citet{ossenkopf92}, which have bulk density
$\rho_{grain}$=3.0~g~cm$^{-3}$.  The grain size distribution is the
standard \citet{mathis77} MRN power-law $n(a) \propto a^{-3.5}$
with 0.001\,\mum\ $\leq a \leq a_{max}$.  In Paper 1 we showed that
micron size grains were required to fit the width of the silicate
feature, and found that the best fits were obtained with
$a_{max}$=3\,\mum\@.  The presence of large grains has been described
in detail in \S5.3 of that paper.  We therefore fix $a_{max}$ to
be 3\,\mum\ in all the new model calculations. Spherical grains,
and Mie scattering are assumed in the models.

The inner radius ($R_{in}$) and the dust density at the inner radius
($\rho_{in}$) are treated as variable input parameters.  We include
a WD, with $R_{WD}$=3 $R_{\earth}$ and $T_{WD}$=60,000~K in the
spatial grid, placed at an appropriate distance from the AGB star
based on the orbital elements of \citet{gromadzki09a} for each
observation date.  The presence of the WD has a local effect,
increasing the dust temperature by about 200\,K in a region less
than about 1~au, but does not impact the overall mid-IR SED, and
is included nonetheless for consistency.  For the AGB star, its
effective temperature, $T_{eff}$ and luminosity are the required
input parameters.  We use $T_{eff}$ and the stellar radius, $R_{\star}$
and use the Stefan-Boltzmann law to calculate the luminosity.

The best-fit model parameters for each observation are obtained in
three stages.  In the first stage, the values for the stellar
effective temperature and radius were determined using dust-free
blackbody models and matching inferred near-IR ($JHKL$) fluxes and
the observed 6.4 and 7.7\,\mum\ fluxes.  Measurements of these
parameters using interferometric observations of R~Aqr \citep{danchi94,
vanbelle96, wittkowski16} provide us with the range of values for
each that we use in the models.  $T_{eff}$ was constrained to lie
in the range 1800--3000~K, and sampled in steps of 100~K\@.  And
$R_{\star}$ was sampled in steps of 0.05~au between 1.6 and 3.0~au.

The $JHKL$ magnitudes appropriate for the observations were obtained
from the historical SAAO data, choosing dates when both the Mira-phase
and V-magnitude were as close as possible to the values during the
mid-IR observations.  The ISO observation occured during the SAAO
campaign, and for that we interpolate the magnitudes from the nearest
observations, before and after.  For SOFIA Cycles 6 and 7, we chose
SAAO dates for which the Mira-phase matched the SOFIA observations,
but the V-magnitudes were adjusted to match the out-of-eclipse
values.  The SAAO observation dates, V-magnitudes, Mira-phases, and
$JHKL$ magnitudes on those dates are listed in Table~\ref{tabjhkl},
and the AGB properties, $T_{eff}$ and $R_{\star}$, in Table~\ref{tabinput},

For the Cycle 6 and 7 data we also ran models using $JHKL$ values
from SAAO observations obtained at times when the V-magnitudes were
the same as during the SOFIA observations, and phases as close as
possible.  The dates of these observations were Oct-04-1983 for
Cycle 6 when the IR brightnesses were about 0.3 magnitude fainter
and January-13-1984 for Cycle 7, when they were about 1.0 magnitude
fainter.  The models predicted significantly lower AGB luminosities
-- 6100\,$L_{\sun}$ and 3400\,$L_{\sun}$ -- and predicted only about
2/3 the observed 6.4\,\mum\ and 7.7\,\mum\ fluxes.

\begin{deluxetable*}{clcccccc}[ht!]
\tablecaption{SAAO Observation Dates, and JHKL Photometry \label{tabjhkl}}
\tablewidth{0pt}
\tablehead{
\colhead{Observation} 
& \colhead{SAAO Obs.}
& \colhead{V\tablenotemark{a}}
& \colhead{$\phi_{Mira}$\tablenotemark{b}}
& \colhead{$J$}
& \colhead{$H$}
& \colhead{$K$} 
& \colhead{$L$} 
}
\startdata
ISO & interpolated & 8.0\tablenotemark{c} & 0.14\tablenotemark{c} & $0.05$ & $-0.97$ & $-1.47$ & $-2.13$ \\
SFC4 & 2001 Nov 27 & 10.7 & 0.36 & $0.62$ & $-0.48$ & $-1.04$ & $-1.62$ \\
SFC5 & 1987 Dec 14 & 8.5 & 0.19 & $-0.06$ & $-1.05$ & $-1.52$ & $-2.18$ \\
SFC6 & 1987 Dec 14 & 8.5 & 0.19 & $-0.06$ & $-1.05$ & $-1.52$ & $-2.18$ \\
SFC7 & 1985 Nov 21 & 9.5 & 0.24 & $0.18$ & $-0.87$ & $-1.40$ & $-2.04$ \\
\enddata
\tablenotetext{a}{Visual magnitude from AAVSO}
\tablenotetext{b}{$\phi_{Mira}=0.0$ (visual maximum) on 2005 October 6, period = 387.0 days}
\tablenotetext{c}{Interpolated to the date of the ISO observation, therefore identical to
the values in Table~\protect\ref{tabobs}}

\end{deluxetable*}

In the second stage, the AGB properties ($T_{eff}$ and $R_{\star}$)
found in the first stage were used (Table~\ref{tabinput}).  Models
were calculated for each observation with inner radius values of
10, 15, 20 and 25~au, with dust added in each case to match the
observed 11.1\,\mum\ flux.  As in Paper 1, we started with $\rho_{in}$
equal to $10^{-19}$~g~cm$^{-3}$, based on the mass loss rates from
\citet{danchi94}, and adjusted this value until visual inspection
of plots with model spectra overlaid on the data showed a good
match.  The initial values of $\rho_{in}$ lay in the range 3--35$\times
10^{-20}$~g~cm$^{-3}$.

\begin{deluxetable*}{cccccc}[ht!]
\tablecaption{Model Input Parameters for AGB Star Properties \label{tabinput}}
\tablewidth{0pt}
\tablehead{
\colhead{Model} 
& \colhead{$R_{\star}$ (au)} 
& \colhead{$T_{eff}$ (K)} 
& \colhead{$L_{\star}$/$L_{\sun}$}
}
\startdata
ISO & 2.35 & 2500 & 9000 \\
SFC4 & 2.30 & 2200 & 5200 \\
SFC5 & 2.55 & 2400 & 9000 \\
SFC6 & 2.55 & 2400 & 9000 \\
SFC7 & 2.40 & 2400 & 8000 \\
\enddata

\end{deluxetable*}

\begin{deluxetable*}{ccccccc}[ht!]
\tablewidth{0pt}
\tablecaption{Model Results \label{tabmodel}}
\tablehead{
\colhead{Model} 
& \colhead{$R_{in}$ (au)} 
& \colhead{$\rho_{in}$\tablenotemark{a}}
& \colhead{$T_{in}$(K)\tablenotemark{b}} 
& \colhead{$\tau_{10}$}
& \colhead{$dM/dt$\tablenotemark{c}}
& \colhead{Reduced $\chi^2$\tablenotemark{d}}         
 }
\startdata
ISO  & 10 & 32.3  & 1080 & 0.12 & 6.4 & 1.8 \\
{}   & 15 & 17.0  &  900 & 0.09 & 7.5 & 1.4 \\
{}   & 20 & 11.3  &  790 & 0.08 & 8.9 & 1.8 \\
{}   & 25 &  8.3  &  710 & 0.07 & 10.1 & 2.9 \\	
{}   & {} & {}    &   {} &  {}  & {}  & {}  \\
SFC4 & 10 & 14.0  &  950 & 0.05 & 2.8 & 0.9 \\
{}   & 15 &  7.5  &  780 & 0.04 & 3.3 & 0.7 \\
{}   & 20 &  5.3  &  670 & 0.04 & 4.1 & 0.7 \\
{}   & 25 &  4.0  &  600 & 0.04 & 4.9 & 0.8 \\		
{}   & {} & {}    &   {} &  {}  & {}  & {}  \\
SFC5 & 10 & 11.3   & 1090 & 0.04 & 2.2 & 0.9 \\
{}   & 15 &  6.0   &  900 & 0.03 & 2.7 & 0.9 \\
{}   & 20 &  4.0   &  770 & 0.03 & 3.2 & 1.1 \\
{}   & 25 &  3.0   &  680 & 0.03 & 3.7 & 1.6 \\	
{}   & {} & {}    &   {} &  {}  & {}  & {}  \\
SFC6 & 10 & 15.5   & 1090 & 0.06 & 3.1 & 0.5 \\
{}   & 15 &  8.0   &  900 & 0.04 & 3.5 & 0.4 \\
{}   & 20 &  5.3   &  780 & 0.04 & 4.1 & 0.4 \\
{}   & 25 &  3.8   &  700 & 0.03 & 4.6 & 0.5 \\
{}   & {} & {}    &   {} &  {}  & {}  & {}  \\
SFC7 & 10 & 13.0   & 1060 & 0.05 & 2.6 & 2.0 \\
{}   & 15 &  7.0   &  860 & 0.04 & 3.1 & 1.5 \\
{}   & 20 &  4.5   &  740 & 0.03 & 3.5 & 1.3 \\
{}   & 25 &  3.3   &  680 & 0.03 & 4.0 & 1.3 \\
\enddata

\tablenotetext{a}{Dust density at $R_{in}$, in units of 10$^{-20}$~g~cm$^{-3}$ for grain bulk density of 3.0\,g~cm$^{-3}$.}
\tablenotetext{b}{Dust temperature at $R_{in}$.}
\tablenotetext{c}{Total mass loss rate in $10^{-7}$\,M$_{\odot}$~yr$^{-1}$; assumes 
spherically symmetric outflow, gas-to-dust mass ratio of 263 to 1, and zero 
drift velocity between the dust and the gas.}
\tablenotetext{d}{$\chi^{2}/(n-1)$ with n = 241, 202, 195, 183, 179 for ISO, SFC4, SFC5, SFC6, SFC7 spectra, respectively.}

\end{deluxetable*}

\begin{figure}[ht!]
\plotone{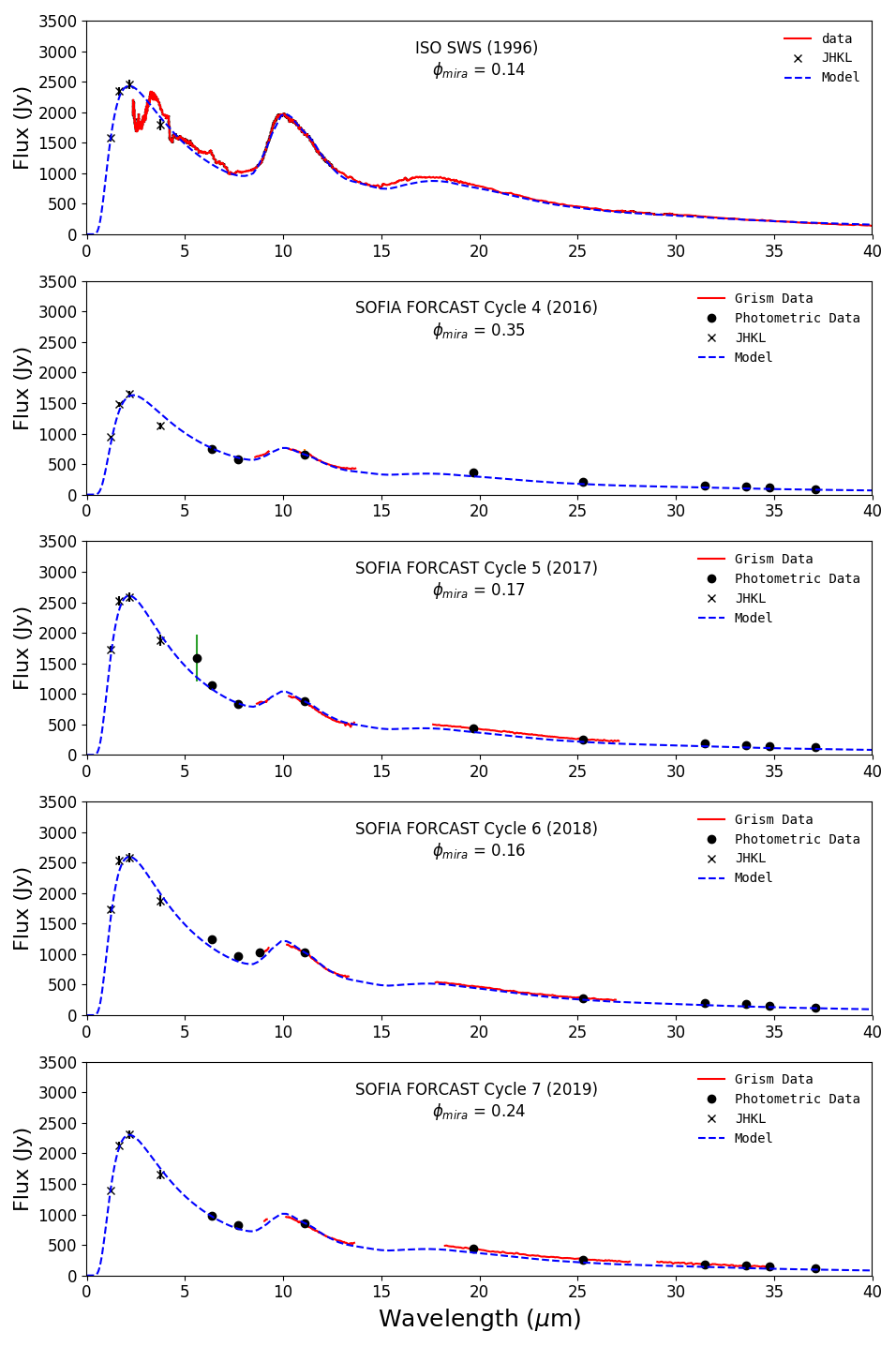}
\caption{Plots of the $R_{in}=15$~au best-fit models overlaid on
the data.  The region between 9.2 and 10.3\,\mum, which is affected
by telluric ozone, has been removed in the SOFIA FORCAST grism
spectra.  The JHKL values used are given in Table~\protect\ref{tabjhkl}.}
\label{figmodel}
\end{figure}

In the third and final stage of the modeling, the stellar temperatures
and radii were held fixed for each observation.  Then, a set of
models were calculated for each of $R_{in}$ = 10, 15, 20 and 25~au,
in each case varying $\rho_{in}$ in steps of $0.25\times
10^{-20}$~g~cm$^{-3}$ around the value found in the second stage.
The Pearson $\chi^2$ statistic was calculated on the difference
between the model and the data as a simple residual sum of squares.
For the FORCAST data, all the photometric points 6.4\,\mum\ and
longer were used as well as the G111 grism data-points, excluding
the region between 9.2 and 10.3\,\mum, which is affected by telluric
ozone.  For the ISO data, the flux values at the FORCAST photometric
points, plus the entire spectral region between 8.4 and 13.7\,\mum\
(corresponding to the G111 coverage) were used.  In each set of
models with fixed $R_{in}$, there was a value of $\rho_{in}$ for
which $\chi^2$ was a clear minimum, which was then selected as the
best-fit model.  The properties of the best-fit models are given
in Table~\ref{tabmodel}.  The reduced-$\chi^2$ is also tabulated,
but we note that the statistic was minimized to find the appropriate
$\rho_{in}$, and not used for estimating the goodness-of-fit.
Figure~\ref{figmodel} shows the synthetic SEDs from the best-fit
models overlaid on the data. For each observation we choose the
model with $R_{in}=15$~au.  In each case, the four best-fit models
for the different values of $R_{in}$ are indistinguishable from
each other in the wavelength range from the NIR through $\approx$18\,\mum,
beyond which they begin to diverge slightly.


\paragraph{New ISO, SFC4, SFC5 Models:} For the first step in our
earlier modeling we did not use these SAAO data, but instead 2MASS
$JHK$ values for ISO and SOFIA Cycle 5, and average values from
\citet{gromadzki09b} for SOFIA Cycle 4.  The input parameters and
the model results have changed.  For ISO and SFC5, the effective
temperatures became 100 and 200~K lower, resectively, and the
luminosity became about 17\% less in both cases.  For SFC4, the
effective temperature became higher by 200~K, and the luminosity
increased by about 27\%.  Consequently the best-fit models' values
for $\rho_{in}$ increased for ISO and SFC5, and decreased for SFC4.
The results, described in \S5 of Paper 1 do not change: (i) within
the one pulsation cycle that includes SFC4 and SFC5, $\rho_{in}$
decreases between phase 0.35 and 0.17, and (ii) there is a substantial
decrease in $\rho_{in}$ between the ISO observations and the SOFIA
observations.  The first result is discussed in greater detail in
\S5, below.

The dust temperatures in each radial zone are determined by RADMC-3D\@.
The values predicted at the inner radius, $T_{in}$, represent the
dust condensation temperatures, and are listed in Table~\ref{tabmodel}.
In all the models, $T_{dust}(r)$ decreases approximately as $r^{-0.5}$.
The values of $T_{in}$ for the best-fit models, 600--1090~K, are
consistent with the range of temperatures at which silicates are
able to form in circumstellar shells \citep[e.g.][]{tielens90,
gobrecht16}.  These values are also consistent with dust temperatures
obtained for R~Aqr in various earlier studies.  \citep{papoular83,
anandrao86, danchi94, lesidaner96, jurkic18}.

RADMC-3D calculates the dust opacity, which depends on the optical
constants, the grain bulk density, and the grain size distribution.
For all our models these parameters remain the same and yield
$\kappa_{10}=2388.3$~cm$^{2}$~g$^{-1}$.  The optical depths at
10\,\mum\ through the dust shell were obtained by integrating
$\kappa_{10}\rho_{dust}(r)$ between $R_{in}$ and $R_{out}$. The
results (Table~\ref{tabmodel}) are consistent with those obtained
for R~Aqr in earlier studies \citep{papoular83, lesidaner96}, as
well as those obtained for the low mass-loss rate AGBs by \citet{suh04}.
These low optical depths are also consistent with the silicate
feature being seen in emission.

The models predict $\rho_{in}$ in the range (3.25--32.25)$\times
10^{-20}$~g~cm$^{-3}$.  Assuming a steady wind speed of 16.7~km~s$^{-1}$
\citep{mayer13}, and a gas-to-dust mass ratio of 263, valid for
silicates \citep{danchi94}, the total predicted mass loss rates
range from $2.2\times10^{-7}$ to $1.0\times10^{-6}$ \,M$_{\odot}$~yr$^{-1}$
(Table~\ref{tabmodel}).  These values are slightly higher than the
recent estimate by \citet{ramstedt18} based on the CO J=3\ra2 line
emission, but much lower than the earlier estimate of \citet{bujarrabal10}
based on the CO J=2\ra1 and J=1\ra0 lines.  Our original estimate
of $\rho_{in}$ was based on values of the mass loss rate and velocity
in \citet{danchi94}, and therefore as expected our model predictions
are consistent with their results.

\section{Discussion}

In Paper 1 we analyzed the variability of the dust emission within
a pulsation cycle by examining the SOFIA Cycle 4 ($\phi_{Mira}$=0.35)
and Cycle 5 ($\phi_{Mira}$=0.17) data, using spherically symmetric
radiative transfer models.  The overall qualitative results presented
there are unchanged, but our new models require a revision of the
parameter values.  The effective temperature of the AGB star increases
from 2200 to 2400~K between Cycle 4 and Cycle 5 observations, and
the stellar radius increases from 2.30 to 2.55~au, which implies a
factor 1.7 increase in the luminosity (Table~\ref{tabinput}).  This
is a smaller contrast than found earlier, where the luminosity
increased by a factor 2.6 between the two epochs.  Consequently,
the implied differences in the dust density at the inner radius are
also smaller than found previously.

Table~\ref{tabmodel} shows several best-fit models for each
observation, each for a different value of $R_{in}$.  There is a
degeneracy induced by these parameters, which cannot be resolved
within our model framework.  For instance, if $R_{in}$ is assumed
to be the same over a pulsation period, then $\rho_{in}$ is about
1.2 to 1.3 times higher at the time of Cycle 4 compared to Cycle 5
observations, and the dust condensation temperature is lower by
about 100~K\@.  If, instead, $T_{in}$ is assumed to be the same,
then the contrast in $\rho_{in}$ is larger.  As a specific case,
we consider the SFC4, $R_{in}$=15\,au and SFC5, $R_{in}$=20\,au
models.  $T_{in}$ is about the same (770\,K) for both models, and
the ratio of $\rho_{in}$ is about 1.9 between SFC4 and SFC5. The
models allow for $\rho_{in}$ to be constant over a pulsation cycle,
which would require that the dust condenses further away from the
AGB star near minimum than near maximum.

The Cycle 6 observations were obtained almost exactly one pulsation
period after the Cycle 5 observations, and show a significant
increase in the mid-IR fluxes.  The FORCAST fluxes are higher by
5--10\% over most of the bandpass, and about 16\% higher at 11.1\,\mum,
which samples the silicate emission feature.  The Cycle 7 observations
were obtained one pulsation cycle after Cycle 6, and at a slightly
later Mira-phase.  Between Cycles 6 and 7, the 11.1\,\mum\ flux has
fallen by about 17\%, and the fluxes at longer wavelengths by about
10\%, bringing them to about the same level as in Cycle 5, and well
above the fluxes in Cycle 4 (Fig.~\ref{figdata}).

The first stage of the model-fitting used the same $JHKL$ values
for both Cycles 5, and 6, and used the 6.6\,\mum\ and 7.4\,\mum\
fluxes in determining the AGB star parameters, which were found to
be the same for both epochs, yielding $T_{eff}=2400$~K and
$L=9000$\,$L_{\odot}$ (Table~\ref{tabinput}).  The best-fit models
predict that for a given value of $R_{in}$, the dust density at the
inner radius is higher by about 30\% for Cycle 6 than for Cycle 5.
The models for Cycle 7 yield the same $T_{eff}$ for the AGB star
as for Cycles 5 and 6, and a slightly smaller radius, implying a
12\% drop in the luminosity (Table~\ref{tabinput}).  The models
predict that for a given $R_{in}$ the dust density, $\rho_{in}$,
for Cycle 7 is about 0.86 the value for Cycle 6.  The degeneracies
described above when comparing Cycle 4 and Cycle 5 models, apply
here as well.  The models do not rule out the case, for instance,
that $R_{in}$ was larger during the Cycle 6 observations than during
Cycle 7, and the dust density lower.

\subsection{Changes in the Dust Emission with Orbital Phase}

The mid-IR flux decreased between the time of the ISO observations,
which took place near apastron, and the SOFIA Cycle 5 observations
just before the onset of the eclipse.  By Cycle 6, the flux has
started increasing again.  In Paper 1 we suggested that the dust
production rate in the AGB star wind reaches a peak near apastron,
when the WD is beyond the dust condensation radius, and drops as
the separation between the two stars decreases and the hot WD
inhibits dust formation.

Within the generally accepted scenario that R~Aqr is now approaching
periastron and the WD is moving between us and the AGB star, the
increase in the mid-IR flux in the Cycle 6 and 7 SOFIA observations
is readily explained in a qualitative way.  It is due to both the
enhancement of the accretion flow rate, and the orbital geometry
which makes this flow and the disk around the WD visible to us.
And, as first suggested by \citet{willson81}, the intervening
material is responsible for the accompanying decrease in visual
brightness.  A persistent dusty disk in the system also explains
the presence of micron size grains, as we noted in Paper 1.

Our models suggest that the above scenario is plausible, but are
not sufficient to demonstrate it quantitatively.  There are two
basic limitations, the first of which is that our models assume
spherically symmetric dust distributions, while the real distribution
is likely to be more disk-like in the regions closest to the stars,
similar to what is seen in gas tracers \citep{ramstedt18, bujarrabal18}.
The second, even more fundamental limitation is one hinted at above,
that both the dust production rate and the accretion rate are being
modulated on orbital time-scales.  The dynamics of the flow needs
to be modeled to obtain the dust distribution at different orbital
phases, and the dust production in the AGB wind in the presence of
a hot WD needs to be modeled to constrain the results of the radiative
transfer calculations.

\subsection{Variation with Mira-Phase}

The sequence of FORCAST observations were obtained over the course
of four successive pulsation cycles of R~Aqr.  Although the onset
of the eclipse midway through the sequence complicates the issue,
we can usefully examine the dependence of various parameters on the
Mira-phase of the system by using the fortuitious circumstance that
the phase was almost exactly the same for Cycles 5 and 6, and
assuming that we can scale Cycle 4 values to Cycle 5, and Cycle 7
values to Cycle 6.  We note that the AGB star parameters were
determined based on out-of-eclipse $JHKL$ values and therefore are
consistent with the above assumption.

\begin{figure}[ht!]
\plotone{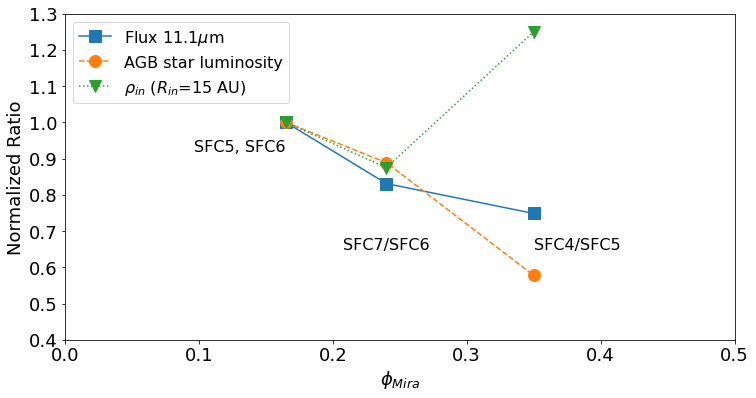}
\caption{The observed 11.1\,\mum\ flux, the model input AGB luminosity,
and the model predicted dust density at the inner radius (for
$R_{in}=15$~au models) plotted against the Mira-phase.  These
quantitites are normalized to 1.0 at phase 0.165, and the ratios
are based on values from Tables~2, 4 and 5, respectively.}
\label{figphase}
\end{figure}

The variation of the 11.1\,\mum\ flux (an observed tracer of the
strength of the silicate emission feature), the AGB star luminosity
(an input to the model), and the dust density at the inner radius
(the key model prediction) with Mira-phase are shown in
Fig.~\ref{figphase}.  These values have been normalized to 1.0 at
$\phi_{Mira}=0.165$ in the plots, which is the average of the Cycle
5 and Cycle 6 phases.  The change between phases 0.16 and 0.24 is
modest, with the AGB luminosity, dust density, and the 11.1\,\mum\
flux all decreasing in lock-step.  Then, by phase 0.35 the AGB
luminosity has fallen to about 65\% of its value at phase 0.24.
This trend in luminosity is consistent with the early finding that
the luminosity maximum for long-period variables is reached at a
pulsation phase of about 0.14 \citep{pettit33}, and confirmed more
recently by \citet{smith06}.  Along with the decrease in the
luminosity, the dust density has increased by over 30\%, between
these phases, and the overall effect is a slight decrease in the
11.1\,\mum\ flux.

Our results provide further evidence of one of the conclusions
presented in Paper 1, that we do not need to invoke any change in
the optical properties of the dust at different Mira phases.  Rather,
we find that the variations in the AGB star temperature and luminosity
are sufficient to explain the changes in the silicate emission over
a pulsation cycle.

\section{Concluding Remarks} 

We have presented new SOFIA/FORCAST observations of R~Aqr, which
were obtained after the onset of the eclipse of the AGB star by its
WD companion and the accretion flow.  The drop in visual magnitude
is accompanied by an increase in the mid-IR flux, and in particular
the strength of the silicate emission feature.  Radiative transfer
models indicate that the underlying stellar spectrum heating the
dust is relatively unaffected by the eclipse, and the increase in
the mid-IR flux is due to higher dust densities in the outflow.

The structure of the outflow and accretion is expected to change
significantly over the course of a binary orbit.  The changing
separation between the WD and the AGB modulates the flow structure,
which in turn influences the dust production efficiency.  The
exploration of non-spherical dust distributions based on realistic
accretion models is the topic of an ongoing study, and will be
presented in a future publication.  Continued monitoring of R~Aqr
in the mid-IR will be crucial for understanding how the dust outflow
and accretion are modulated by the binary orbit.


\acknowledgments We thank the SOFIA SMO staff for their contribution
towards obtaining and calibrating the data for the community. This
work was supported in part by SOFIA GO grants SOF 06-0005 and SOF
07-0097 to the Space Science Institute.  PAW acknowledges a research
grant from the South African National Research Foundation. We are
grateful to Fred Marang and Francois Van Wyk who made most of the
previously unpublished $JHKL$ observations.

\facilities{SOFIA, ISO, AAVSO}


\appendix

\section{Near-infrared Photometry}
\restartappendixnumbering 
\startlongtable
\begin{deluxetable}{lRRRR}
\tablecaption{Infrared Photometry\label{tabIRphot}}
\tablehead{
\colhead{JD$-$2440000}&\colhead{$J$}&\colhead{$H$}&\colhead{$K$}&\colhead{$L$}\\
days & \multicolumn{4}{c}{(mag)}
}
\startdata
2592\tablenotemark{a}  & 2.08 & 0.79  &-0.07  &-1.17  \\
2620  & 2.13 & 0.87  &-0.02  &-1.17  \\
2646  & 2.14 & 0.87  &-0.04  &-1.24  \\
2681  & 2.13 & 0.88  &-0.04  &-1.33  \\
2704  & 2.23 & 0.95  & 0.01  &-1.34  \\
2737  & 2.13 & 0.86  &-0.10  &-1.47  \\
2940  & 2.75 & 1.43  & 0.34  &-1.03  \\
2971  & 3.17 & 1.76  & 0.64  &-0.80  \\
2995  & 3.48 & 2.01  & 0.86  &-0.57  \\
3050  & 3.19 & 1.79  & 0.62  &-0.97  \\
3079  & 3.19 & 1.86  & 0.59  &-0.98  \\
3109  & 3.19 & 1.83  & 0.60  &-1.01  \\
3136  & 2.85 & 1.51  & 0.26  &-1.29  \\
3326.68 & 2.33 & 1.05  & 0.13  &-1.13  \\
3347.59 & 2.51 & 1.04  & 0.15  &-0.99  \\
3387.54 & 2.62 & 1.23  & 0.34  &-0.83  \\
3452.32 & 2.36 & 1.08  & 0.18  &-1.03  \\
4448.53 & 0.60 & -0.56 & -1.13 &-1.92  \\
4579.30 & 0.85 & -0.22 & -0.83 &-1.68  \\
4582.32 & 0.79 & -0.25 & -0.86 &-1.69  \\
4616.27 & 0.63 & -0.37 & -0.96 &-1.80  \\
4769.67 & 0.17& -0.86& -1.36& -2.28 \\ 
4772.62 & 0.19& -0.82& -1.35& -2.25 \\ 
4775.61 & 0.28& -0.79& -1.33& -2.19 \\ 
4800.59 & 0.33& -0.74&     & -2.12 \\ 
4815.58 & 0.42& -0.52& -1.01& -2.05 \\ 
4819.57 & 0.44& -0.61& -1.18& -2.01 \\ 
4852.44 & 0.78& -0.32& -0.91& -1.73 \\ 
4901.34 & 1.15& 0.03 &-0.60 &-1.45  \\
4904.31 & 1.15& 0.04 &-0.60 &-1.46  \\
4907.36 & 1.25& 0.07  &-0.63  &-1.43   \\
4912.32 & 1.16 &  &  &-1.46   \\
4914.28 & 1.20 & 0.11  &-0.48  &-1.35   \\
4915.39 & 1.18 &   &   &-1.48   \\
4918.33 & 1.23 & 0.07  &-0.57  &-1.43   \\
4947.28 & 1.06& -0.02& -0.70& -1.56 \\ 
4955.30 & 1.08&  0.02&-0.61 &-1.57  \\
4958.29 & 1.03& -0.02& -0.64& -1.51 \\ 
4960.28 & 0.87& -0.14& -0.78& -1.66  \\ 
5157.64 & 0.21& -0.82& -1.35& -2.16 \\ 
5181.54 & 0.26& -0.78& -1.33& -2.08 \\
5193.52 & 0.34& -0.72& -1.28& -2.01 \\
5215.49 & 0.49& -0.57& -1.12& -1.94 \\ 
5251.36 & 0.74& -0.35& -0.92& -1.70 \\ 
5280.31 & 0.88& -0.21& -0.78& -1.56 \\ 
5543.53 & 0.00& -1.02& -1.56& -2.35 \\ 
5544.58 & 0.02& -0.99& -1.50& -2.35 \\ 
5545.52 & 0.02& -1.02& -1.55& -2.36 \\ 
5546.55 & 0.05& -0.95& -1.47& -2.27 \\ 
5548.56 & 0.01& -0.99& -1.49& -2.32 \\ 
5567.42 & 0.03& -0.98& -1.56& -2.26 \\ 
5571.52 & 0.06& -0.96& -1.48& -2.30 \\ 
5595.42 & 0.24& -0.73& -1.22& -2.11 \\ 
5602.49 & 0.26& -0.74& -1.31& -2.08 \\ 
5612.38 & 0.32& -0.70& -1.26& -2.03 \\ 
5613.46 & 0.39& -0.62& -1.19& -1.98 \\ 
5615.41 & 0.42& -0.63& -1.20& -1.97 \\ 
5626.34 & 0.54& -0.50& -1.07& -1.93 \\ 
5637.43 & 0.64& -0.39& -0.97& -1.86 \\ 
5649.32 & 0.79& -0.28& -0.87& -1.77 \\ 
5681.31 & 1.10& 0.02 &-0.65 &-1.57  \\
5688.29 & 1.16& 0.11 &-0.59 &-1.52  \\
5701.26 & 1.23& 0.14 &-0.52 &-1.45  \\
5713.26 & 1.25& 0.17 &-0.53 &-1.48  \\
5873.68 & 0.32& -0.61& -1.17& -2.18 \\ 
5892.65 & 0.28& -0.70& -1.24& -2.11 \\ 
5896.63 & 0.25& -0.70& -1.24& -2.15 \\ 
5920.58 & 0.16& -0.82& -1.34& -2.20 \\ 
5958.43 & 0.12& -0.89& -1.41&   \\
6023.31 & 0.54& -0.46& -1.03& -1.81 \\ 
6031.31 & 0.63& -0.38& -0.96& -1.80 \\ 
6070.30 & 1.05& -0.01& -0.65& -1.50 \\ 
6075.29 & 1.07&  0.03& -0.62& -1.44 \\ 
6222.66 & 0.68& -0.21&  & -1.81 \\ 
6303.49 & 0.08& -0.95& -1.44& -2.22 \\ 
6351.39 & &  &  & -2.23 \\ 
6391.30 & 0.18& -0.87& -1.40& -2.04 \\ 
6596.67 & 0.73& -0.23& -0.80& -1.67 \\ 
6638.60 & 0.34& -0.58& -1.09& -1.96 \\ 
6658.60 & 0.18& -0.76& -1.27& -2.07 \\ 
6692.46 & 0.02& -0.94& -1.43& -2.16 \\ 
6712.40 &-0.02& -1.01& -1.48& -2.21 \\ 
6747.40 & 0.04& -0.97& -1.46& -2.15 \\ 
6753.32 & 0.08& -0.94& -1.43& -2.10 \\ 
6780.34 & 0.22& -0.79& -1.31& -1.94 \\ 
6792.29 & 0.32& -0.71& -1.21& -1.95 \\ 
7014.57 & 0.42& -0.48& -1.00& -1.80 \\
7056.48 & 0.10& -0.84& -1.32& -2.08 \\ 
7113.39 &-0.09& -1.07& -1.55& -2.21 \\ 
7144.28 &-0.05& -1.05& -1.52& -2.18 \\ 
7367.60 & 0.77& -0.25& -0.81& -1.56 \\ 
7376.58 & 0.81& -0.20& -0.79& -1.56 \\ 
7393.51 & 0.79& -0.18& -0.75& -1.56 \\ 
7394.55 & 0.78& -0.20& -0.76& -1.59 \\ 
7447.34 & 0.22& -0.70& -1.18& -1.96 \\ 
7685.69 & 1.06& -0.05& -0.68& -1.39 \\ 
7732.68 & 0.77& -0.23& -0.80& -1.51 \\
7745.57 & 0.75& -0.23& -0.80& -1.62 \\
7759.53 & 0.79& -0.21& -0.78& -1.60 \\
7779.52 & 0.80& -0.18& -0.77& -1.60 \\
7821.43 & 0.44& -0.52& -1.03& -1.85 \\
7840.27 & 0.24& -0.69& -1.18& -1.97 \\
8073.63 & 1.05& -0.05& -0.68& -1.43 \\
8086.70 & 1.02& -0.06& -0.68& -1.48 \\
8129.56 & 0.80& -0.17& -0.78& -1.69 \\
8165.48 & 0.72& -0.23& -0.79& -1.63 \\
8213.32 & 0.39& -0.54& -1.07& -1.90 \\
8224.37 & 0.34& -0.63& -1.14& -1.97 \\
9146.69 & &  &  & -1.81 \\
9212.63 & 0.86& -0.25& -0.84& -1.48 \\
9497.67 & 0.13& -0.89& -1.39& -2.03 \\
9581.59 & 0.91& -0.17& -0.77& -1.44 \\
9614.55 & 0.95& -0.13& -0.71& -1.41 \\
9637.27 & 0.82& -0.20& -0.80& -1.48 \\
\enddata
\tablenotetext{a}{no exact times available for the first 13 observations }
\end{deluxetable}


\end{document}